
\input phyzzx.tex
\vskip-10pt
\hfill
{July, 1993\par}
\vskip-10pt
\hfill
{KHTP-93-08\par}
\vskip-15pt
\hfill
{SNUTP-93-38\par}
\title{\seventeenbf Possible Tests of Conformal Turbulence through Boundaries}
\author{B.K. Chung, Soonkeon Nam,
\foot{E-mail address: nam@nms.kyunghee.ac.kr}
Q-Han Park,
\foot{E-mail address: qpark@nms.kyunghee.ac.kr}
and H.J. Shin
\foot{E-mail address: shin@SHIN.kyunghee.ac.kr} }
\address{\it  Department of Physics\break
	   and\break
       Research Institute for Basic Sciences\break
	  Kyung Hee University\break
	  Seoul, 130-701, Korea}
\abstract{
We investigate various boundary conditions in two dimensional turbulence
systematically in the context of conformal field theory.
Keeping the conformal invariance, we can either change the shape
of boundaries through finite conformal transformations, or insert
boundary operators so as to handle more general cases.
Effects of such operations will be reflected in physically measurable
quantities such as the energy power
spectrum $E(k)$ or the average velocity profiles.
We propose that these effects can be used as a possible test of
conformal turbulence in an experimental setting.
We also study the periodic boundary
conditions, i.e. turbulence on a torus geometry.
The dependence of moduli parameter $q$ appears explictly
in the one point functions in the theory, which can also be tested.
}
\endpage


\REF\turb{See for example,
L.D. Landau and E.M. Lifshitz, {\it Fluid Mechanics}, Pergamon, Oxford (1984).}
\REF\book{A.S. Monin and A.M. Yaglom, {\it Statistical Fluid
Mechanics}, The MIT Press, Cambridge (1975).}
\REF\Kol{A.N. Kolmogorov, J. Fluid Mech. {\bf 13} (1962) 82.}
\REF\Kraichnan{R. Kraichnan, Phys. Fluid, {\bf 10} (1967) 1417.}
\REF\turbexp{B. Legras, P. Santangelo, and R. Benzi, Europhys. Lett.
{\bf 5} (1988) 37.}
\REF\PolyI{A.M. Polyakov, Princeton preprint PUPT-1341,
hep-th/9209046.}
\REF\PolyII{A.M. Polyakov, Nucl. Phys. {\bf B396} (1993) 367.}
\REF\Matsuo{Y. Matsuo, Mod. Phys. Lett. {\bf A8} (1993) 619.}
\REF\Falkovich{G. Falkovich and A. Hanany, WIS-92-88-PH, hep-th/9212015;
WIS-93-5-PH, hep-th/9301030.}
\REF\Low{D.A. Lowe, Mod. Phys. Lett. {\bf A8}
(1993) 923.} \REF\Ferretti{G. Ferretti and Z. Yang, Europhys. Lett. {\bf 22}
(1993) 639.}
\REF\khturbii{B.K. Chung. S. Nam, Q-H. Park, and H.J. Shin,
Kyung Hee Univ. preprint, KHTP-93-07, hep-th/9307091 .}
\REF\CardyII{J.L. Cardy, Nucl. Phys. {\bf B324} (1989) 581.}
\REF\CardyI{J.L. Cardy, Nucl. Phys. {\bf B240} [FS12] (1984) 514.}
\REF\khturb{B.K. Chung, S. Nam, Q-H. Park, and H.J. Shin,
Phys. Lett. {\bf 309B} (1993) 58.}
\REF\Burkhardt{T.W. Burkhardt and T. Xue, Phys. Rev. Lett. {\bf 66}
(1991) 895; Nucl. Phys. {\bf B354} (1991) 653.}
\REF\Callan{C.G. Callan, C. Lovelace, C.R. Nappi, and S.A. Yost, Phys. Lett.
{\bf 206B} (1988) 41; Nucl. Phys. {\bf B308} (1988) 221.}
\REF\Cappelli{A. Cappelli, C. Itzykson, and J.B. Zuber, Nucl. Phys.
{\bf B280} (1987) 445.}
\REF\Sonoda{H. Sonoda, Nucl. Phys. {\bf B311} (1988/89) 417.}
\REF\Bagger{J. Bagger, D. Nemeschansky, and J.B. Zuber, Phys. Lett.
{\bf 216B} (1989) 320.}
\REF\IsingI{A.E. Ferdinand and M.E. Fisher, Phys. Rev. {\bf 185} (1969) 832.}
\REF\IsingII{P. Kleban, G. Akinci, R. Hentschke, and K.R. Brownstein,
 J. Phys. A {\bf 19} (1986) 437.}

Turbulence is a ubiquitous phenomenon in the physics of fluids,
with scales ranging  from the astronomical to centimeters.\refmark\turb\
In the statistical approach to turbulence,\refmark\book\ Kolmogorov's theory
of the inertial range provides a simple but powerful tool in
obtaining the power spectra of turbulence.
In three dimensions,
there is an energy cascade from large scale to the smaller ones.
The similarity argument gives rise to
the well known $k^{-5/3}$ law\refmark\Kol\ which
has successfully described some experimental data.
However, in two dimensions, there is a fundamental difference,
i.e. there is the vorticity conservation along the fluid
particle trajectories.  This leads to an enstrophy cascade towards small
scales, giving
the $k^{-3}$ law.\refmark\Kraichnan\
Although various numerical simulations\refmark\turbexp\
suggest different power spectra, usually with values less than $-3$,
theoretical explanation for these values so far has been quite elusive.

Polyakov's recent work on the theory of two dimensional turbulence
shed a new light on the problem.\refmark{\PolyI, \PolyII }
In this approach, the stream function $\psi$
is assumed to be a primary field of a minimal conformal field theory (CFT)
with conformal dimension $\Delta_\psi$.
By solving the Hopf equation and
the constant enstrophy flux condition, he proposed that many non-unitary
CFT's become exact solutions of turbulence.
It was also suggested that a specific boundary matching condition
could provide a selection rule among embarrassingly many solutions obtained by
Polyakov
and others.\refmark{\Matsuo - \Ferretti }
Moreover, due to the nature of non-unitary conformal theory,
boundary conditions are essential in defining a theory
itself.\refmark\khturbii\
On the other hand, real experimental or numerical tests require definite
boundary
 conditions, either periodic or non-periodic. Therefore a proper understanding
of
boundary conditions in conformal turbulence remains as an important open
problem.

The purpose of this letter is to investigate various boundary conditions
systematically in the context of CFT approach to turbulence.
We show that physically measurable quantities, such
as energy spectrum or velocity profiles, are affected by specific boundary
conditions
 and we compute them in case of large momentum $k$.
Keeping the conformal invariance, we can either change the shape
of the boundary, through finite conformal transformations, or insert
boundary operators in the boundary.\refmark\CardyII\
We also study the periodic boundary
condition, i.e. conformal turbulence restricted on a torus.
In this case we obtain the dependence of moduli parameter $q$
through the one point functions in the theory.
The boundary effects on the energy spectrum and the velocity profiles, and also
the $q$-dependence of the one point functions, can be tested experimentally.
We propose these tests as a possible verification of conformal turbulence.

A prototypical CFT on a manifold with  boundary
is $(p,p')$ minimal model in the upper half plane,
$\hbox{Im} z =y >0$,\refmark\CardyI\  with the primary fields
$\Phi_{(r,s)}$ of conformal
dimensions $\Delta_{rs}=[(ps-p'r)^{2}-(p-p')^{2}]/4pp'$.
In the above, $p$ and $p'$ are co-prime positive integers and
$0<r<p,\ 0<s<p'$.
In Ref.[15], we have considered how the energy spectrum is affected by the
boundary condition, and in Ref.[12], we have solved the Hopf equation with
the constant enstrophy condition for such a model, and the simplest solution
was
the $(2,33)$ minimal model with $\Phi_{(1,10)}$ of conformal dimension
$\Delta_{1,10}=-3$ as the stream function.
In this case the energy spectrum turns out to
be,\refmark{\khturb, \khturbii }
$$E_{(x,y)}(k)=\sum_{i}C^{\phi_{i}}_{\psi\psi}\ k^{4\Delta_{\psi}-2
\Delta_{\phi_{i}}+1}\langle \phi_{i}(z)\rangle_{b}+ {\hbox
{secondaries}}, \eqno(1)$$
where $\phi_{i}$ are primary operators of conformal dimension
$\Delta_{\phi_{i}}$, and $C^{\phi_{i}}_{\psi\psi}$ are structure constants
of operator product expansion of two $\psi$'s.
As long as the one point function of $\phi_{i}$ does not vanish, the
energy spectrum away from the boundary is dominated by
$k^{4\Delta_{\psi}-2\Delta_{\phi}+1}$
where $\phi$ is the least dimension operator among $\phi_i$'s in Eq.(1).
In the (2,33) model, $\phi $ is $\Phi_{(1,17)}$,
and the exponent of $k$ is $ -41/11\sim -3.72723$.
The one point function evaluated in the upper half plane is such that
$$\langle \phi_{i}(z)\rangle_{b}=d_{\phi_{i}}y^{-2\Delta_{\phi_{i}}},\eqno(2)$$
where $d_{\phi_{i}}$ are arbitrary constants, which are parameters of the
theory which gain physical meaning through the velocity correlations.
The velocity one point functions are easily obtained from the one
point function of stream function by differentiation:
$$\langle v_{x}(x,y)\rangle_{b}=-2\Delta_{\psi}d_{\psi} y^{-2\Delta_{\psi}-1},
\qquad\langle v_{y}(x,y)\rangle_{b}=0.\eqno(3)$$
We see that there is no average velocity normal to the boundary, but only
the average flow along the boundary. Such a velocity profile is to be
expected from a rigid and slippery boundary.

In turbulence problem, one is often interested in various shapes of boundaries.
One of the advantage of conformal turbulence is that it
 is rather easy to get different boundaries simply by performing finite
conformal transformations.
Under a finite conformal transform; $z \rightarrow f(z),\ \bar{z}
\rightarrow \bar{f}(\bar{z})$, a primary
field transforms as
$$ \phi'(z',\bar{z}')=\left({df\over dz}\right)^{-\Delta_{\phi}}
\left({d{\bar f}\over d{\bar z}}\right)^
{-\overline{\Delta}_{\phi}}\phi(z,\bar{z}).\eqno(4)$$
{}From this, one can immediately evaluate one point functions of primary
fields and the energy spectrum from eqs.(1) and (2).

First, we consider the case of a strip of width $L$ which requires
the transform;
$t= x_{1}+ix_{2}= {L\over i\pi}\ln z$, $0<x_1<L, \ \ -\infty < x_2 < \infty.$
Using eq.(4), we get the one point function of the stream function;
$$\langle \psi (t)\rangle _{\hbox{strip}} =
\left({\pi\over L}\right)^{2\Delta_{\psi}} {d_{\psi}\over
\left(\sin{\pi\over L} x_{1}\right)^{2\Delta_{\psi}}}.\eqno(5)$$
which gives rise to the average velocity;
$$\eqalign{\langle v_{1}\rangle_{\hbox{strip}} &=   0,\cr
\langle v_{2}\rangle_{\hbox{strip}} &= 2 d_{\psi}\Delta_{\psi}
\left({\pi\over L}\right)^{2\Delta_{\psi}+1}
{\cos \left({\pi\over L}x_{1}\right)
\over \sin\left({\pi\over L} x_{1} \right)^{2\Delta_{\psi}+1}}.}
\eqno(6)$$
Note that we have an antisymmetric velocity profile across the $x_1=L/2$ line.
Later on, we will see that more general type of velocity profiles are possible
if we consider the insertion of boundary operators.
For other primary fields, we have similar expressions for the one point
function, and thus the energy spectum can be obtained using eq.(1):
$$
E_{(x,y)}(k)=\sum_{i} C^{\phi_{i}}_{\psi\psi}
k^{4\Delta_{\psi}-2\Delta_{\phi_{i}}+1}
\left({\pi\over L}\right)^{2\Delta_{\phi_{i}}} {d_{\phi_{i}}\over
\left(\sin{\pi\over L} x_{1}\right)^{2\Delta_{\phi_{i}}}}+\cdots.\eqno(7)$$

For large $L$, there will be a region formed around the center
of the strip in which the one point function of
the least dimension operator dominates the energy
power spectrum. This region may be identified with the usual
inertial range of Kolmogorov where small scale fluctuations become
isotropic despite of the large scale anisotropy.

Using a further conformal transformation;
$$w={\exp[i\pi ({t\over L}-{1\over 2})]-1\over
\exp[i\pi ({t\over L}-{1\over 2})]+1}, \eqno(8)$$
we can map a strip onto the interior of a circle
of radius one.
Evaluating the one point function of the stream function, we get
$$\langle \psi(w)\rangle_{\hbox{circle}}= -{4^{\Delta_{\psi}}d_{\psi}\over
(1-r^{2})^{2\Delta_{\psi}}},\eqno(9)$$
where $w=r e^{i\theta}$.
The velocity profile is then
$$\langle v_{r} \rangle_{\hbox{circle}}=0, \ \langle
v_{\theta}\rangle_{\hbox{circle}}
={4^{\Delta_{\psi}+1}d_{\psi} r\over (1-r^{2})^{2\Delta_{\psi}+1}}.
\eqno(10)$$
This certainly gives the velocity profile that circles around
the origin, and displays the full circular symmetry.
Of course, we can make yet further conformal transformations;
$f(z)=z+a^{2}z^{-1}$ to map the circle into an ellipse,
or $f(z)=U[(z+\delta)+b^{2}(z+\delta)^{-1}]$ to map into the symmetrical
Zhukovski aerofoil where the energy spectrum and the velocity profiles can be
obtained straightforwardly.

In order to get more general velocity profiles, we need to impose different
boundary
conditions  of 2-d CFT on the upper and lower parts of boundary in case of a
strip.
In general, imposing an arbitrary boundary condition will break the conformal
invariance. However there are certain cases where  we can change boundary
values
but keeping the conformal invariance.
These cases arise when we insert a boundary operator $\phi_{b}$,
which itself is a primary or a secondary conformal field, in the boundary of
upper half plane,
and perform a finite conformal transform to the strip.
Such a boundary operator has already been considered in the surface critical
phenomena\refmark\Burkhardt\ and in the open string theory.\refmark\Callan\

Operationally, the effect of a boundary operator $\phi_{b} (0)$ inserted
at $z=0$ is to change the boundary condition of the half plane such that
boundary values become discontinuous at
the origin. Since the global conformal transformation requires a point added at
infinity, we need
a conjugate boundary operator at infinity to take care of the boundary change
at infinity.
Thus, when we evaluate a one point function of a conformal field in the
upper half plane, together with the image point in the lower half
plane and the two boundary operators inserted,
we need to evaluate a 4-point function in the plane
geometry.\refmark\Burkhardt\
In general, an $n$-point function in the upper half plane with an inserted
boundary
operator satisfies a differential equation for a $2n+2$ point function for the
bulk case.
The differential equations for the correlation functions follow from
the conformal Ward identity, which specifies the change in the
correlation function under an arbitray infinitesimal coordinate
transformation.
For an operator degenerate at the level two, there is a second-order
differential equation in terms of which the bulk four point function reduces
to the one point function in the upper half plane with a boundary operator.

Following the above prescription, we get  the most general form of one point
function,
$$\langle \phi_{i}(z,{\bar z})\rangle_{\phi_{b}} = y^{-2\Delta_{\phi_{i}}}
{\cal F}(\cos\theta),\eqno(11)
$$
where $z=x+iy$, $\cos\theta =x/|z|$ and $\langle \ \rangle_{\phi_{b}}$ denotes
the expectation
on a half plane with the insertion of an operator $\phi_{b}$ at the origin.
In general, the explicit form of $\cal F$ is difficult to obtain except
for the cases where the underlying conformal field theories  are degenerate
at the level two. Nevertheless Eq.(11) in its own form is already quite
suggestive.
When we fix the angle $\theta$ and move away from the boundary, we
have only the $y$-dependence, and also through eq.(1), this behavior is
reflected in the
energy spectrum.
A finite conformal trasform of Eq.(11) once again leads to the one point
function on a strip;
$$ \langle \phi_{i}(t) \rangle_{\hbox{strip}}
= d_{\phi_{i}}\left({\pi \over L}\right)^{2\Delta_{\phi_{i}}}
{{\cal F}(
\cos {\pi \over L}x_{1}) \over (\sin {\pi \over L}x_{1})^{2\Delta_{\phi_{i}}}}
\eqno(12)$$
which in particular shows that the velocity profile is no longer antisymmetric.

Let us be more specific on the form of ${\cal F}$. Even though we can not find
an explicit form of ${\cal F}$ in general, the property of ${\cal F}$ under the
modular transformation restricts further the form of ${\cal F}$ in the
following way; in the presense of boundary operators which juxtapose different
boundary conditions, there is a relation\refmark\CardyII\ between the boundary
values of one point function,i.e. the boundary value on the right hand side of
the boundary operator,
$x>0$, and that on the left hand side $x<0$. These are related through the $S$
matrix of the modular
transformation $\tau\rightarrow -\tau^{-1}$ with the modular parameter
$\tau$:
$$\chi_{rs}(q)=\sum_{r',s'}S_{(r,s)}^{(r',s')}\chi_{r's'}(\tilde{q}),\eqno(13)$$
where $q=e^{2\pi i\tau}$ and $\tilde{q}=e^{-2\pi i/\tau}$, and
the Virasoro characters are given by
$\chi_{rs}(q)=q^{-{c\over 24}}\sum_{n=0}^{\infty}d_{rs}(n)q^{n},$
with $d_{rs}(n)$ giving number of independent states at level $n$.
The sum is over integers $r'$ and $s'$ for the ranges $1\leq r'\leq p'-1$ and
$1\leq s' \leq p-1$, with the condition $s'p'\leq r'p$.
The relation, specifically for the boundary operator $\phi_{r',s'}$ and for the
one point function of $\phi=\phi_{r,s}$ in eq.(11) is given by
$$ {{\cal F}(\cos 0)\over {\cal F}(\cos \pi)}=
{ S_{(r,s)}^{\ (r',s')}S_{(1,1)}^{\ (1,1)}\over
S_{(r,s)}^{\ (1,1)}S_{(1,1)}^{\ (r',s')}},\eqno(14)$$
where the $S$ matrix for $(p,p')$ representation of the
Virasoro algebra has components such that\refmark\Cappelli\
$$S_{(r,s)}^{\ (r',s')}= 2 \sqrt {2\over pp'}
(-1)^{rs'+r's+1}\sin\left({\pi p r r'\over p'}
\right)\sin\left({\pi p' s s'\over p}\right).\eqno(15)$$

As for a concrete example, we consider the $p=2, p'=5$ minimal CFT solution
of two dimensional turbulence.
Since there is only one primary field apart from the identity,
there is only one  nontrivial boundary operator,  the stream function itself.
The $S$ matrix in this case is given as follows:
$$S=\left(\matrix{
-{2\sin(2\pi/5)\over\sqrt{5}} & -{2\sin(6\pi/5)\over\sqrt{5}} \cr
-{2\sin(6\pi/5)\over\sqrt{5}} & -{2\sin(18\pi/5)\over\sqrt{5}} \cr}\right),
\eqno(16)$$
where the first column or row corresponds to the identity operator
and the second to the primary field $\phi_{1,2}$.
The  one point function of the stream function is
$$\langle \psi (z)\rangle _{\psi}=
y^{2/5} {\cal F} (\cos \theta),\eqno(17)$$
where ${\cal F}(\cos \theta)$ is given in terms of a linear
combination of hypergeometric functions;
$${\cal F}(\cos\theta)=A\ {}_2 F_{1}\left
({2\over 5},{1\over 5};{1\over2};\cos^{2}\theta\right)
+B\cos\theta\ {}_2 F_{1}\left({9\over10},{7\over 10};
{3\over 2};\cos^{2}\theta\right),\eqno(18)$$
for some constants $A$ and $B$.
{}From the relation as in  Eq.(14), we have
$${{\cal F}(\cos \pi)\over{\cal F}(\cos 0)}
= {\sin({2\pi\over 5})\sin({18\pi\over 5})
\over \sin {}^{2}({6\pi\over 5})}.\eqno(19)$$
which determine $A={\Gamma(11/10)\Gamma(1/2)
\over \Gamma(9/10)\Gamma(7/10)} $ and
$B={\Gamma(11/10)\Gamma(-1/2)\over \Gamma(2/5)\Gamma(1/5)}$ so that
we can write $\cal F(\cos\theta)$ in a closed form;
$${\cal F}(\cos\theta)\sim {}_{2}F_{1}\left({4\over5},{2\over 5};
{11\over10};{1-\cos\theta\over 2}\right).\eqno(20)$$
Putting this back into Eq.(11) gives the exact one point function of the stream
function, and the velocity profile through differentiation.
The energy spectrum is now given by Eq.(1) as a function of the distance
from the boundary operator as well as the angle formed by the boundary
and the line joining the origin and the point of evaluation of the
energy spectrum.

Finally, we consider the periodic boundary case,i.e. conformal turbulence
on a torus. In general, the multipoint functions of the minimal model on a
torus can be evaluated in terms of  the generalized Virasoro characters
$\chi_{\lambda}(q,z)$ where $q=\exp (2\pi i\tau), \ \ (\hbox{Im}\tau>0).$
In the pinching limit,  $q\rightarrow 0$, the torus degenerates
into a long cylinder and we can expand the one point function on a torus
in $q$ which  becomes
a sum of three point functions
on the plane.\refmark{\Sonoda,\Bagger }
Explicitly, this takes the form,
$$\langle \phi_{i}(z,\bar{z}) \rangle _{\hbox{torus}}\sim
{1\over Z(q)} (q\overline q)^{-{c\over 24}}\sum_{l}
q^{\Delta_{l}}\bar{q}^{\bar{\Delta}_{l}}
\langle \phi_{l}|\phi_{i}(1,1)|\phi_{l}\rangle_{\hbox{sphere}}
+\cdots\eqno(21)$$
where the sum is over the primary and secondary fields of the theory, and
$Z(q)$ is the partition function, which itself can be expanded in
$q$. The three point functions on a sphere can be determined uniquely by the
conformal symmetry up to coefficients of OPE. We note that the $q$-dependence
of the one point function
arises in powers of $q$ with the conformal dimensions as power exponents.
Here, we have a good example of the case where the one point function
of a conformal field does not vanish and moreover even the value itself
is determined, unlike the upper half plane case.
On a torus, one point function of a primary field is independent of the
position $z$ while it is explicitly dependent on $q$. This leads to the
vanishing  of the average velocity while the energy spectrum becomes dependent
on $q$ through one point functions in Eq.(1). The shape dependence through
periodic boundary conditions has already appeared in the study of critical
Ising model in a
rectangular geometry,\refmark\IsingI\
and conformal field theory prediction has been
tested.\refmark\IsingII\
So, it would be a real test of conformal
turbulence if we observe boundary effects predicted as in this paper in the
laboratory or in a computer simulation.

\noindent
\ack{
We would like to thank Je-Young Choi for discussions.
This work was supported in part by the program of
Basic Science Research, Ministry of Education,
and by Korea Science and Engineering Foundation,
and partly through CTP/SNU. QP thanks K.Kang and Physics Department of Brown
University for their support through SNU-CTP exchange program during his visit.
}
\singlespace
\vfill\eject
\refout
\end